\documentclass[preprint,aps,floats,subeqn,showpacs,amssymb]{revtex4}
\usepackage{epsfig}

\newcommand{\be}{\begin{equation}}
\newcommand{\ee}{\end{equation}}
\newcommand{\bea}{\begin{eqnarray}}
\newcommand{\eea}{\end{eqnarray}}
\newcommand{\bml}{\begin{mathletters}}
\newcommand{\eml}{\end{mathletters}}

\begin{document}

\tighten

\draft




\title{Black strings and solitons in five dimensional space-time
 with positive cosmological constant}
\renewcommand{\thefootnote}{\fnsymbol{footnote}}
\author{Y. Brihaye and T. Delsate \footnote{yves.brihaye@umh.ac.be}  }
\affiliation{Faculte des Sciences, Universite de Mons-Hainaut, 7000 Mons, Belgium}

\date{\today}
\setlength{\footnotesep}{0.5\footnotesep}

\begin{abstract}
We consider the classical equations of the Einstein-Yang-Mills
model in five space-time dimensions and in the presence of a cosmological constant.
We assume that the fields do not depend on the extra dimension and that they are
spherically symmetric with respect to the three standard space dimensions.
The  equations are then transformed into
a set of ordinary differential equations that we solve numerically.
We construct new types of regular (resp. black holes) solutions
which, close to the origin (resp. the event horizon) resemble
the 4-dimensional gravitating monopole (resp. non abelian black hole)
but exhibit an unexpected asymptotic behaviour.
\end{abstract}

\pacs{04.20.Jb, 04.40.Nr}
 \maketitle
\section{Introduction}

In the last years, there has been an increasing
attention to space-times involving more than four dimensions
and particularly to brane-world models \cite{brane} to describe space, 
time and matter.
These assume the standard model fields to be
confined on a 3-brane embedded in a higher dimensional manifold.
A large number of higher dimensional black holes has been studied in
recent years. The first solutions that have been constructed
are the hyperspherical generalisations of well-known black holes
solutions such as the Schwarzschild and Reissner-Nordstr\"om solutions
in more than four dimensions \cite{tan} as well
as the higher dimensional Kerr solutions \cite{mp}.
In $d$ dimensions, these solutions have horizon topology $S^{d-2}$.

However, in contrast to 4 dimensions black holes with different horizon
topologies should be possible in higher dimensions. 
An example is a 4-dimensional
Schwarzschild black hole extended trivially
into one extra dimension, a so-called
Schwarzschild black string. These solutions have been discussed extensively
especially with view to their stability \cite{gl}.
A second example, which is important due to its implications for
uniqueness conjectures for black holes in higher dimensions, is the black ring
solution in 5 dimensions with horizon topology $S^2\times S^1$ \cite{er}.
 
The by far largest number of higher dimensional
black hole solutions constructed so far are solutions of the
vacuum Einstein equations, respectively Einstein-Maxwell equations.

On the other hand, it is believed that topological defects have occured and played a role
during some phase tansitions in the evolution of the Universe, see e.g.
\cite{vilenkin}.
In particular, magnetic monopoles
\cite{thooft} must have been produced during the GUT
symmetry breaking phase transition. The actual non-observation of magnetic
monopoles
leads to constraints which have to be implemented into the models of inflation.
On the other hand observational evidence obtained in the last years
\cite{astrocc} favours the possibility that space-time has an accelerated
expansion which could be related to a positive cosmological constant.

It is therefore natural to examine the properties of the various
topological defects in presence of a cosmological constant, or said
in other words, in asymptotically DeSitter space-time.
Recently \cite{bhrds} the magnetic monopole and the sphalerons
occuring in an SU(2)
gauge theory spontaneously broken by a scalar potential were constructed
in an asymptotically DeSitter space-time and it was found that the 
asymptotic decay of the matter field is not compatible with a finite mass.

The first example of higher dimensional
black hole solutions containing non-abelian
gauge fields have been discussed in \cite{bcht}. These are non-abelian
black holes solutions of a generalised 5-dimensional Einstein-Yang-Mills
system with horizon topology $S^3$. 
Using ideas of \cite{volkov,bh1},
SU(2)-black strings with $S_2\times S_1$
topology were constructed in \cite{hartmann}.
Several regular and black hole solutions of an Einstein-Yang-Mills
model have been constructed recently with different symmetries
\cite{bhr,bh2,bch}.
These solutions are non-abelian black hole solutions in 3+1-dimensions
extended to one extra dimension.

In \cite{bbh,faissal} the Einstein-Yang-Mills model
in five dimensions with gauge group SU(2) was considered with a positive
cosmological constant.
The metric and gauge fields were assumed to be independent
of the extra dimension and chosen to be spherically
symmetric in the standard three space-like dimensions.
By adopting a Schwarzschild-dilaton type  parametrisation
for the metric, it was found that the equations can be
integrated only up to a maximal value of the radial coordinate,
say $r=r_c$.
A coordinate singularity occurs at $r=r_{c}$.
In this paper, we adopt the parametrisation of the metric
used in \cite{mrs} and we show that the solution of \cite{faissal}
can be extended up to spatial infinity in these coordinates.

We give the model and  the two parametrisation of the metric
in Sect. II.  The relevant reduced action for the gravitating
and matter parts are presented in Sect. III together with the
boundary conditions. The numerical results
corresponding to solutions regular at the origin and solutions
possessing an event horizon
are discussed 
in  IV. The summary is given in Section V.
\section{The Model and the ansatz}

The Einstein-Yang-Mills Lagrangian
in $d=(4+1)$ dimensions is
given by:
\begin{equation}
\label{action}
  S = \int \Biggl(
    \frac{1}{16 \pi G_{5}} (R - 2 \Lambda_5 ) 
    - \frac{1}{4 e^2}F^a_{M N}F^{a M N}
  \Biggr) \sqrt{g^{(5)}} d^{5} x
\end{equation}
with the SU(2) Yang-Mills field strengths
$F^a_{M N} = \partial_M A^a_N -
 \partial_N A^a_M + \epsilon_{a b c}  A^b_M A^c_N$
, the gauge index
 $a=1,2,3$  and the space-time index
 $M=0,...,5$. $G_{5}$, $\Lambda_5$ and $e$ denote
respectively the $5$-dimensional Newton's  and cosmological
constants and the coupling
constant of the gauge field theory. $G_{5}$ is related to the Planck mass
$M_{pl}$ by $G_{5}=M_{pl}^{-3}$ and $e^2$ has the dimension of 
$[{\rm length}]$.

In this paper, we assume that the metric and the matter fields
are independent on the extra coordinate $y$ and we will use 
a spherically symmetric ansatz for the fields.

Our aim is to construct non-abelian regular and black strings
solutions  which are
spherically symmetric in the four-dimensional space-time
and are extended
into one extra dimension. The topology of these non-abelian
black strings will thus be $S^2\times \mathbb{R}$ or
 $S^2\times S^1$ if   the extra coordinate
$y$ is chosen to be periodic.

We will use two different coordinates systems for the metric.
On the one hand, the metric can be parametrized according to \cite{volkov}
as follows
\begin{equation}
\label{metric1}
g^{(5)}_{MN}dx^M dx^N =
e^{-\xi}\left[-A^{2}Ndt^2+N^{-1}dr^2+r^2 d\theta^2+r^2\sin^2\theta
d^2\varphi\right]
+e^{2\xi} dy^2   \ \ : \ \ {\rm type (1) }
\label{metric}
\  \end{equation}
where $N,A, \xi$ are function of the coordinate $r$ only. 
For the gauge fields, we use the spherically symmetric ansatz
\cite{thooft} :
\begin{equation}
\label{matter}
{A_r}^a={A_t}^a=0 \  , \   {A_{\theta}}^a= (1-K(r)) {e_{\varphi}}^a
\ , \
{A_{\varphi}}^a=- (1-K(r))\sin\theta {e_{\theta}}^a   \  , \
{\Phi}^a=v H(r) {e_r}^a \
\end{equation}
where $v$ is a mass scale.

The classical equations corresponding to the model
above were studied in \cite{bbh} and more recently in \cite{faissal}
with the appropriate
boundary conditions corresponding to regular solution at the
origin $r=0$ and black string solutions presenting a regular event
horizon on a cylinder, i.e. with $N(r_h)=0$.
The equations were solved numerically and it was found
that the solutions having the regular behaviour at $r=0$ or $r=r_h$
exist only up to a maximal value $r=r_{c}$ (with the
value $r_{c}$ depending on both $\alpha$ and $\Lambda$).
For $r \to r_{c}$ the fields behave according to
\begin{equation}
     N(r) \sim  N_c(r_c-r) \ \ , \ \
\xi(r) \sim \xi_i +  \xi_c \sqrt{ (r_c-r)}   \ \ , \ \
    A(r) = A_c (r_c-r)^{-a}
\end{equation}
where $N_c, \xi_i,  \xi_c, A_c, a$ are constants (with $a > 0$)
  depending on $\alpha$ and $\Lambda$.
The behaviour of the function $\xi$ clearly indicates an absence of
analyticity at $r=r_c$.

In view of these diffculties, we have tried to study the equation
by using a different parametrisation of the metric.
For instance, we use the length element used e.g. in \cite{mrs}.
It reads
\begin{equation}
ds^2 = -b(x)dt^2 + \frac{dx^2}{f(x)} +
g(x) \left( d\theta^2 + \sin^2(\theta)d\varphi^2 \right) +
a(x)dy^2         \ \ : \ \ {\rm type (2) }
\label{metric2}
\end{equation}
where the radial variable is named $x$ (to avoid confusion
with $r$ used in (\ref{metric1})). The arbitary redefinition of the 
coordinate $x$ is left unfixed at this stage
but on it will be fixed later according to $g(x)=x^2$. The ansatz
for the matter fields is identical to (\ref{matter}) apart from the
fact that the functions $K,H$ now depend on $x$.
Using $g(x)=x^2$ as gauge fixing, the correspondence between the two sets
of functions in (\ref{metric1}) and (\ref{metric2}) reads~:
\be
\label{conversion}
x     = r e^{-\xi(r)/2} \ \ , \ \
f(x) = N (1 - \frac{r}{2} \frac{d \xi}{d r})^2 \ \ ,  \ \
a(x) = e^{2\xi(r)} \ \  , \ \
b(x) = A(r)^2N(r)e^{-\xi(r)}
\ee
\section{Equations of motion}
Using an appropriate rescaling of the radial variable $ e v x \to x $
the classical equations associated
to the action (\ref{action}) lead to a set of ordinary differential 
equations  depending
on the fundamental coupling $\alpha \equiv 4\pi \sqrt{G_{5}} v$
and on the reduced cosmological constant
$\Lambda \equiv 2 \alpha^2 \Lambda_5$.
In the case of the  parametrisation (\ref{metric1}) , the equations
are written in details in \cite{faissal} together with the appropriate
boundary conditions. In the case of the parametrisation (\ref{metric2}),
the equations are lengthy. They can be obtained easily for the
reduced action 
\be
   S_{red}  =  \sqrt{-\tilde g} \frac{1}{\sin \theta} R  + S_{mat}  
   \ \ , \ \ \sqrt{-\tilde g} = g \sin \theta  \sqrt{ab/f}
\ee
with the gravitating part proportional to the Ricci scalar
\be
    \sqrt{-\tilde g} \frac{1}{\sin \theta} 
    R =  - \biggl(\sqrt{abf} (g(\frac{b'}{b}+\frac{a'}{a})+2 g' )\biggr)'
     + g \sqrt{\frac{ab}{f}}\biggl(\frac{f}{2} (\frac{g'}{g})^2
     + f \frac{g'}{g}(\frac{a'}{a} + \frac{b'}{b}) + \frac{2}{g}
      + \frac{f}{2} \frac{a' b'}{a b}\biggr)
\ee
and
\be
   S_{mat} = \sqrt{-\tilde g} \biggl(
   \frac{K'^2}{fg} + \frac{1}{2 g^2} (1-K^2)^2 + \frac{1}{2} \frac{H'^2}{f}
   + \frac{1}{g}K^2 H^2
\biggr)
\ee
where the prime denote derivative with respect to 
the rescaled radial variable $x$.
\subsection{Boundary conditions}
In the type 2 system of coordinates, it should be noticed
that the function $a$ and $b$ can be rescaled arbitrarily.
The regularity conditions for a regular solution
 at the origin read  \cite{mrs}
\be
f(0) = 1 \ \ , \ \ a(0) = 1 \ \ , \ \ b(0)=1 \ \ , \ \ b'(0)=0   \ \
\ee
while  black strings possessing an horizon at $x=x_h$ should have
\be
   f(x_h) = 0 \ \ , a(x_h) = 1 \ \ , \ \ b(x_h)=0 \ \ , \ \ b'(x_h) = 1
\ee
in both case a natural choice of the normalisation of $a,b$ has
been supplemented.
For the matter fields,
the boundary conditions for a regular solution at the origin and
the usual asymptotic conditions read
\begin{equation}
\label{bc0}
  \ \ K(0) = 1 \ \ , \ \ H(0)=0 \ \ 
  , \ \ K(\infty)=0 \ \ , \ \ H(\infty)=1 \ \ ,
\end{equation}
In the case of black holes, the regularity at the horizon
imposes some peculiar relations between the values $H(x_h), K(x_h)$
and of their derivatives. These expressions are cumbersome and will
not be presented here.

\subsection{Kasner-type solutions}
The study of the classical equations in the vacuum case,
i.e. for $K(x)=1, H(x)=0$ is interesting by itself. A complete
analysis of black strings in the case of a negative cosmological
constant is reported in \cite{mrs}.
In that paper, solutions were constructed numerically which grow
asymptotically according to
\be
\label{ads}
a(x) = \frac{x^2}{\ell^2} \ \ , \ \ b(x) = \frac{x^2}{\ell^2}
\ \ , \ \ f(x) = \frac{x^2}{\ell^2} \ \ 
{\rm with}  \ \ \Lambda \equiv -\frac{6}{\ell^2}
\ee
(the next terms of these expansions can be found in \cite{mrs}).
Obviously, solutions with the corresponding asymptotic expansion
can be expected in the present context of a positive cosmological
constant.

For both cases (i.e. $\Lambda < 0$ and  $\Lambda > 0$) however,
we obtained another type of asymptotic solutions. This reads
\be
\label{kasner}
      a = x^{2A}  \  , \  b = x^{2B} \  , \  f = f_0 x^{2F} \ \
      {\rm with} \ \   A = B = -2 -  \sqrt{3} \ , \ F = 3 + 2 \sqrt{3}
\ee
and where $f_0$ is a constant. The power-dependence of the metric functions
on the radial coordinate is reminiscent of the well known Kasner solution
in four-dimensional space-time.  It can be checked easily that the Ricci
scalar corresponding to these solutions vanishes asymptotically.
\section{Numerical results}
\subsection{ Regular solutions}
We solved the equations corresponding to the lagrangian (\ref{action})
by numerical methods for several values of $\alpha$ and  $\Lambda > 0$.
In the  system of coordinate (\ref{metric2}),
our results strongly suggest that the solutions approaching the
regular boundary condition at the origin can be extrapolated
 for $x\to \infty$ and that, in this asymptotic limit,
 they approach  a Kasner-type solution (\ref{kasner}).
This is illustrated on Fig. 1 for $\alpha=1, \Lambda =0.0005$.
Close to the origin, these solutions are similar to the gravitating
dilatonic monopole but contrary to our expectation they do not extrapolate
 to a DeSitter space-time for $\Lambda>0$. 
 It seems that the presence of the cosmological
 constant and the corresponding Liouville potential leads to an asymptotic
 space-time obeying the power law (\ref{kasner}) asymptotically.

In order to confirm this result, we also considered the equations in the case
$\alpha = 0$ where the matter fields trivialize by means of $K=1,H=0$. 
The equations therefore
correspond  to 5-dimensional gravity in the presence of a positive cosmological
constant. 
The profiles of the functions $a,b,f$ are  represented in the subplot of Fig. 3.
Comparing Figs.1 and 3, we can appreciate the 
significant deformation of the functions $a,b,f$
due to the matter fields in the region $x \sim 0$.

The double logaritmic scale used on Fig. 3 
demonstrates  the power decay of the metric functions. Our numerical results
strongly indicate that
 $a(x)=b(x)$ in this case. This relation does not hold true for black strings 
 (see next section) but it suggests
 that the solution could be expressed in an explicit form 
 (although we failed to find it so far).
To finish this section, let us mention that, in the case of a 
negative cosmological constant,
a solution regular at the origin also exists; 
it was constructed in \cite{mrs}. In contrast
to the present case, the solution of \cite{mrs}
approaches asymptotically the solution (\ref{ads}).

\subsection{Black hole solutions}
The numerical construction of black hole solutions of the
lagrangian (\ref{action})
is more difficult with the type-2 parametrisation of the 
metric than with the type-1
parametrisation because two functions (for instance $f$ and $b$) vanish at the
event horizon $x_h$, leading to several singular terms in the equations 
(with the type-1 parametrisation, we have $N(r_h)=0$ only).
Nevertheless, the results of \cite{faissal}
suggest that such solutions exist at least locally. Moreover, the
solutions constructed with the type-1 coordinate  provide 
suitable starting profiles for the different functions.
It is worth noticing that, once converted into the type-2 coordinate
system by means of (\ref{conversion}) the  functions $a,b,f$ and their 
derivatives turn out to be smooth
in the neighbourhood of the maximal value $x_c =  r_c e^{-\xi(r_c)/2}$.
Solving the equations
for the type-2 coordinates confirms indeed  the non abelian black
string of \cite{faissal} and further shows that these solutions
can be extended for $x \in [x_h,\infty]$ with $\Lambda > 0$. 

The profile for such a solution
is presented on Fig. 2 for $\alpha=1, x_h=0.3$ and $\Lambda = 0.0005$.
Similarly to the case of regular solutions our numerical results strongly
suggest  that these black hole solution extrapolate between the condition of a regular black string with horizon at
$x=x_h$ and the Kasner-type behaviour (\ref{kasner}) 
for $x\to \infty$.

We also solved the equations in the pure gravity case 
(i.e. $\alpha=0$ and $W=1,H=0$) and obtained
the black string solutions with positive cosmological constant.
These are represented on Fig. 3 (main figure). 
These solutions are the counterparts for $\Lambda > 0$
of the black string solutions presented in \cite{mrs} for $\Lambda <0$.
Let us point out that for $\Lambda<0$ the black string  
solutions extrapolate between a regular horizon
and ADS space-time.
In the neighbourhood of the event horizon $x_h$ the
two solutions look quit similar but they deviate considerably
from each other for $x >> 1$.

Let us finally mention that, integrating the equations 
from $x=\infty$ with (\ref{ads})
as initial condition and with $\Lambda >0$  leads 
to configurations which become singular for $x\to 0$.
A systematic study of black string with 
$\Lambda > 0$ and $d>4$ will be presented elswhere \cite{brs}.

\section{Summary}
The construction of solitons and black string solutions 
for the Einstein-Yang-Mills equations
in a five-dimensional space-time and in the presence 
of a cosmological constant turns out to be
numerically difficult. The problem was adressed 
in \cite{faissal} but it appeared that the system
of coordinates used was not satisfactory, leading
to a coordinate singularity at some maximal value
of the radial coordinate. Here we reconsidered the 
equation with an ansatz of the line element
inspired from the literature about black strings (see e.g. \cite{mrs}).
It turns out that, with the new
coordinates, the solutions can be continued up to $x= \infty$
and our numerical results suggest that
the metric is of the Kasner-type asymptotically. This feature 
seems to hold true for pure gravity
black strings as well as for the non-abelian case.
In the limit $x_h\to 0$ these two types of black strings
approach a non trivial solution which is regular at the origin.
\\
{\bf\large Acknowledgements} 
\\
Y. B. thanks  the
Belgian FNRS for financial support
 and gratefully acknowledges discussions with Eugen Radu.

\newpage
\begin{figure}
\centering
\epsfysize=12cm
\mbox{\epsffile{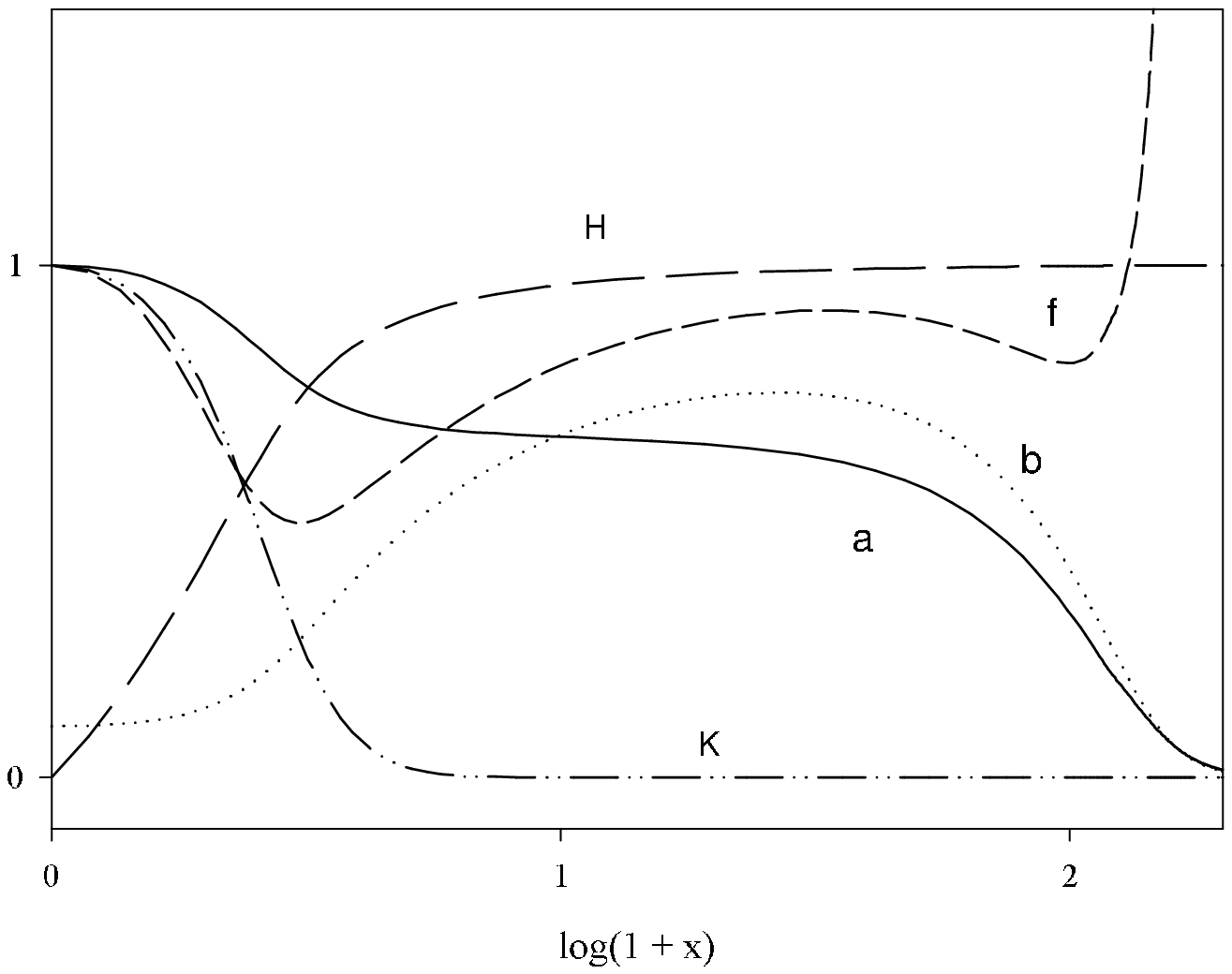}}
\caption{\label{fig1}
The profile for a non-abelian soliton
corresponding to $\alpha=1$, $\Lambda = 0.0005$ }
\end{figure}
\newpage
\begin{figure}
\centering
\epsfysize=12cm
\mbox{\epsffile{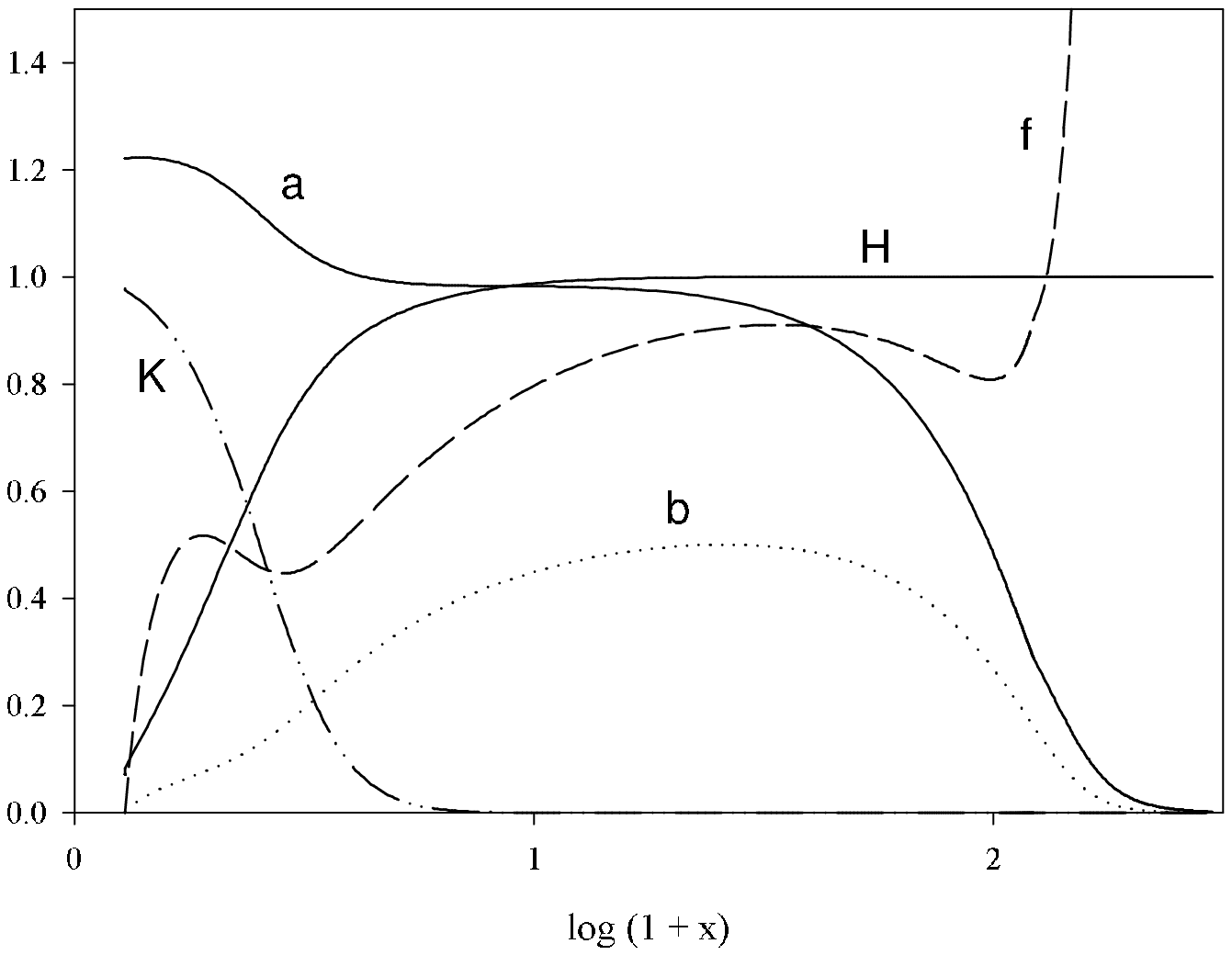}}
\caption{\label{fig1}
The profile for a non-abelian black string corresponding to
$\alpha=1$, $\Lambda = 0.0005$ and $x_h=0.3$ }
\end{figure}

\newpage
\begin{figure}
\centering
\epsfysize=18cm
\mbox{\epsffile{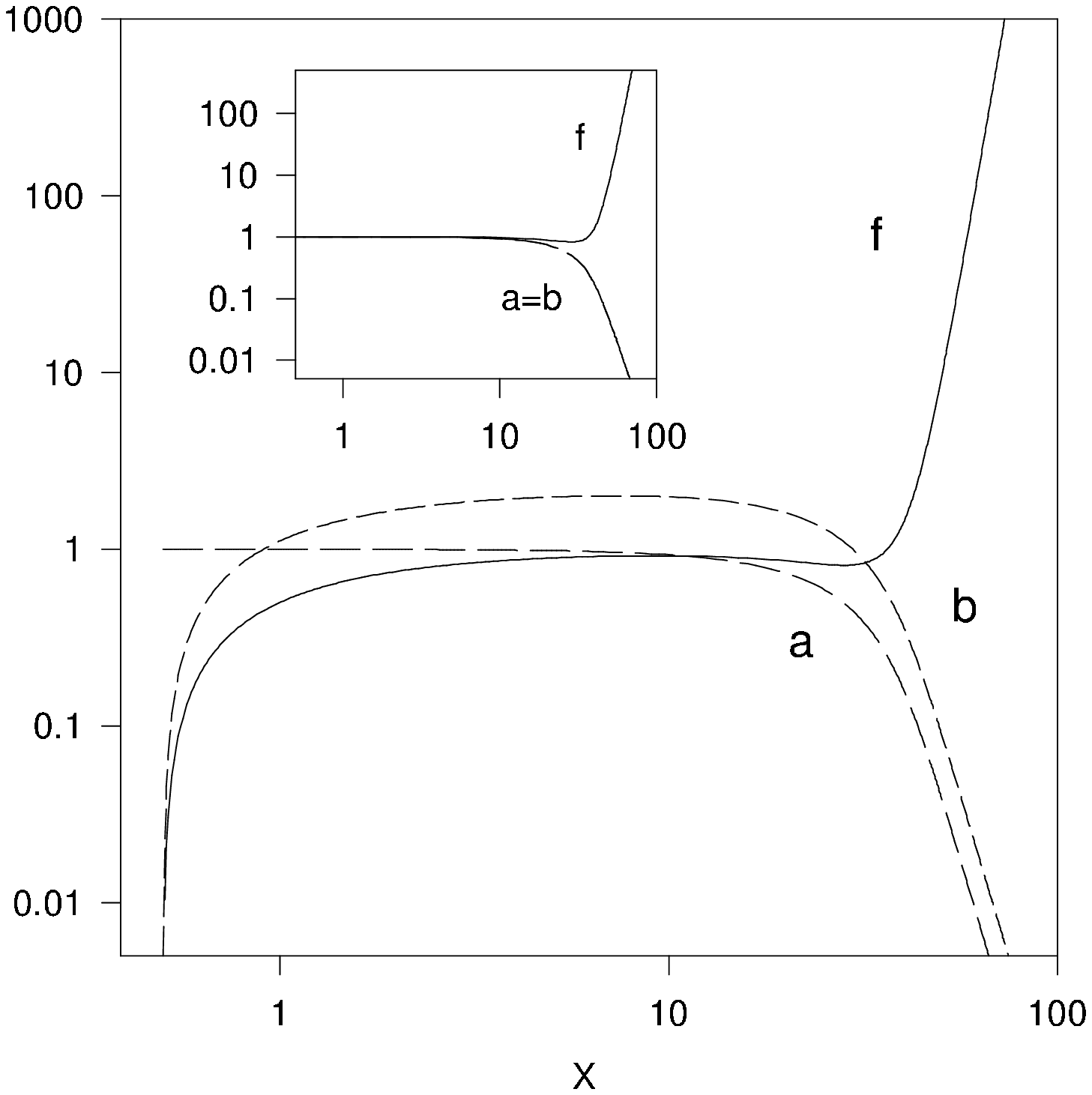}}
\caption{\label{fig1}
The profiles for the pure gravity black string and of the regular solution in the window.}
\end{figure}
\end{document}